**Title:** Lattice nano-ripples revealed in peptide microcrystals by scanning electron nanodiffraction

**Authors:** Marcus Gallagher-Jones[1], Colin Ophus[2], Karen C. Bustillo[2], David R. Boyer[1], Ouliana Panova[2,3], Calina Glynn[1], Chih-Te Zee[1], Jim Ciston[2], Kevin Canton Mancia[1], Andrew M. Minor[2,3], and Jose A. Rodriguez[1]

**Affiliations:**
[1] Department of Chemistry and Biochemistry, UCLA-DOE Institute for Genomics and Proteomics, University of California Los Angeles, Los Angeles, CA 90095, USA.
[2] National Center for Electron Microscopy, Molecular Foundry, Lawrence Berkeley National Laboratory, CA, USA.
[3] Department of Materials Science and Engineering, University of California Berkeley, CA, USA.

**Abstract:** Changes in lattice structure across sub-regions of protein crystals are challenging to assess when relying on whole crystal measurements. Because of this difficulty, macromolecular structure determination from protein micro and nano crystals requires assumptions of bulk crystallinity and domain block substructure. To evaluate the fidelity of these assumptions in protein nanocrystals we map lattice structure across micron size areas of cryogenically preserved three-dimensional peptide crystals using a nano-focused electron beam. This approach produces diffraction from as few as 1,500 molecules in a crystal, is sensitive to crystal thickness and three-dimensional lattice orientation. Real-space maps reconstructed from unsupervised classification of diffraction patterns across a crystal reveal regions of crystal order/disorder and three-dimensional lattice reorientation on a 20nm scale. The lattice nano-ripples observed in micron-sized macromolecular crystals provide a direct view of their plasticity. Knowledge of these features is a first step to understanding crystalline macromolecular self-assembly and improving the determination of structures from protein nano and microcrystals from single or serial crystal diffraction.

**Introduction**

The physical and chemical properties of a protein crystal depend in part on its underlying lattice structure. Changes in the packing of macromolecules within crystals perturb this structure as is exemplified by crystal polymorphism[1]. Packing rearrangements can also lead to deterioration of lattice order and limit the usability of a crystal for structural determination[2,3]. Imperfections in protein crystals can in part be described by the mosaic block model[2–4], in which monolithic crystal blocks or domains tile to form a macro-crystal but vary in size, orientation, and/or cell dimensions[3]. Because directly measuring mosaicity in protein crystals is inherently challenging[2], crystallographic software must estimate disparities in domain block size, shape, and orientation per crystal[5–7], for full and partial Bragg reflections[5,8]. Mosaicity then varies by crystal and is affected by crystal size[9], crystal manipulation[10], and parameters for data collection[11]. The challenge in accurately assessing these models in protein nanocrystals has been highlighted by analysis of diffraction measured using x-ray free electron lasers[6,12].

Direct views of a protein crystal lattice can be obtained by high resolution electron microscopy[13,14]; facilitated by advances in high-resolution imaging[13–17] and cryogenic sample handling techniques[13,18,19]. Cryo electron microscopy (cryoEM) also reveals crystal self-assembly[20–24] and, for two-dimensional protein crystals[21], shows natural variation between unit cells[22,23]. Domain blocks can be identified in cryoEM images of three dimensional lysozyme microcrystals[20], where Fourier filtering helped estimate the location and span of multiple blocks across a single crystal[20].

Macromolecular structures can be obtained from similar nanocrystals by selected area electron diffraction-based methods such as MicroED[25] and rotation electron diffraction[26]. Structures determined by MicroED or similar approaches range in size from small molecules to proteins, including a variety of peptides[27–34]. In MicroED, frozen-hydrated nanocrystals are unidirectionally rotated while being illuminated by an electron beam to produce diffraction movies[35]. These movies are processed by standard crystallographic software[36], and structures are determined and refined using electron scattering factors[37,38]. Structures determined by MicroED are atomically accurate[27,28,32] and mirror those obtained by microfocus x-ray crystallography[33,39]. These structures represent an average over entire crystals or large crystal areas.

In electron microscopy, greater control over illuminated areas is achieved by scanning transmission electron microscopy (STEM), which scans a focused electron beam (typically <1nm) across a sample to produce images of relatively thick bio specimens[40] over large fields of view[41,42]. A variety of sample properties can be probed by collecting electrons from different angular ranges, such as annular dark field detection with low and high scattering angle detectors (ADF, HAADF)[43,44], annular bright field detection (ABF)[45], and differential phase contrast detection[46], providing access to different contrast mechanisms underlying these modalities. These approaches typically rely on monolithic detectors that integrate electrons over a specific angular range originating from the sample at each probe position and attribute the signal intensity to a point on sample[44]. These techniques have been successful in the 3D mapping of atomic features within imperfect crystals of radiation hard material[47].

In contrast, a scanning nanobeam diffraction experiment records diffraction patterns on a two-dimensional pixelated detector at each scan position across a sample. Each $Scan_x$ x $Scan_y$ position of the scan has a $K_x$ x $K_y$ dimension in reciprocal space (diffraction image) resulting in a four-dimensional data set (4DSTEM)[48–50]. This data can then be processed to reconstruct a real space image of the sample corresponding to specific features in the measured diffraction patterns from each scan point, resulting in a greater flexibility in the imaging contrasts obtainable from a single experiment. Cooling of sensitive semi-crystalline organic polymers allows their investigation by 4DSTEM to reveal lattice orientation within nano-thin films[41,42].

Here we analyse beam-sensitive three-dimensional peptide nanocrystals at liquid nitrogen temperatures by 4DSTEM. Our findings address a lack of direct estimates of nano-scale lattice variation in single protein crystals by other diffraction methodologies and their relationship to diffraction data quality. Our measurements reveal effects that may influence MicroED experiments and other nanocrystallography methods including those performed at x-ray free electron lasers[6,12,51–54], or synchrotron-based micro and nanocrystallography[9,55,56].

**Results**

*4DSTEM of three-dimensional beam sensitive peptide nanocrystals*
To assess nanoscale lattice changes within single peptide crystals, we generated two-dimensional maps of their lattice structure by 4DSTEM (Fig 1.). We evaluated nanocrystals of a prion peptide segment with sequence QYNNQNNFV, whose structure has been previously determined (PDB ID 6AXZ) by MicroED to sub-Å resolution[33]. Crystals analysed by 4DSTEM were equivalent in shape and size to those evaluated by MicroED; most were needle shaped and several microns in their longest dimension, but less than a micron thick and wide (Supplementary Figure 1.). Crystals that lay over holes on the quantifoil® grid were found to produce the highest contrast signal and were chosen for analysis by 4DSTEM.

During 4DSTEM data collection, an approximately six nanometre electron beam (Supplementary Figure 2.) was scanned coarsely across a grid while regions of interest were identified by low-mag low-dose STEM imaging. Fine scans were performed on up to three micrometre long regions of single nanocrystals while diffraction patterns were measured at each step using a direct electron detector (Fig 1c.). Beam parameters in 4DSTEM scans were comparable to those of prior studies (supplementary table 1), with an estimated fluence of ~1 e$^-$/Å$^2$ in regions of the crystal exposed to the beam. For all 4DSTEM scan points we applied the lowest dose necessary to detect Bragg reflections (Fig. 1d,e, Supplementary Figure 1.). We confirmed that the dose was sufficiently low by measuring several overlapping scans from the same area of a single crystal and found that decay in the highest resolution reflections occurred only after multiple scans of the same area (Supplementary Figure 3.). The probe dwell time and diffraction image exposure were synchronized and produced full-frame patterns at a rate of 400 per second (Supplementary Figure 4.).

These parameters yielded diffraction patterns whose resolution extended beyond 1.4 Å (Supplementary Figure 1.), measured from only an estimated 1,000-15,000 diffracting molecules given an effective illumination area of 6 by 6 nm, a crystal thickness of 100-500 nm and unit cell constants of a=4.94, b=10.34, c=31.15, alpha=94.21, beta=92.36, gamma=102.2. Bragg diffraction was observed only from within the bounds of the crystals, as identified by annular dark field STEM images (Fig. 1, Supplementary Figure 1, 3). Diffraction patterns integrated over micron-sized regions of single crystals, matched MicroED patterns measured from a micron-sized selected area (Supplementary Figure 1). Diffraction in 4DSTEM patterns was minimally occluded by the angular dark field detector (Fig. 1a).

*Diffraction pattern reconstruction by hybrid electron counting*
Conventional electron counting in HRTEM is achieved by thresholding images acquired using a fast direct electron detector based on estimates of detector readout and Landau noise[17]. Coincidence loss is minimized by operating the detector at a sufficiently high frame rate and low electron dose[17]. This procedure limits the dynamic range in diffraction images because coincidence loss is difficult to escape at Bragg reflections, where signal is concentrated compared to background regions (Supplementary Figure 4). This phenomenon is exacerbated most by detection of the focused central beam, where coincidence loss is high (Supplementary Figure 5). To benefit from the improved signal to noise afforded by electron counting, whilst maintaining some of the dynamic range lost due to coincidence, we implemented a hybrid method for signal detection (Supplementary Figure 5). This procedure introduced little change in the signal observed at most Bragg reflections compared to standard counting, but improved estimates of central beam intensity (Supplementary Figure 5).

Raw patterns collected on a K2 direct electron detector at 400 frames per second were processed using a workflow that involves background subtraction, normalization, thresholding, and pixel-level stepwise counting (Supplementary Figure 4).

The hybrid counting approach was implemented as follows: We used the distribution of pixel intensities within an entire 4DSTEM scan to estimate background signal and categorize measured pixel intensities into counting bins corresponding to single or multiple electron counts (Supplementary Figure 4e, 5a). Based on this criterion, the majority of pixels in a pattern were zero valued (Supplementary Figure 4e); a small fraction (< 1%) within the regions where Bragg reflections were expected received single counts and a lesser minority of pixels were assigned multiple counts corresponding to measurement of coincident electrons (Supplementary Figure 5a.). Of the latter group, most events occurred within the region illuminated by the central (000) focused beam disc (Supplementary Figure 5a (insets)), particularly over thin sample regions or holes in the carbon support (Supplementary Figure 5b,c). The range of signal counts in processed patterns spanned values from 1-35 but very few pixels were assigned values greater than 10 (Supplementary Figure 5a). Whilst the true degree of coincidence cannot be measured by this method, comparison of these patterns to those processed with a binary threshold shows an increase in dynamic range achieved by hybrid counting protocols (Supplementary Figure 4f, 5b,c).

Pixel-wise processing of data also allowed us to correct diffraction pattern shifts due to beam scanning using centre of mass alignment (Supplementary Figure 6.). The majority of scans presented minimal shifts; some scans showed preferential drift of the diffraction pattern along a single orientation (Supplementary Figure 6c,d (inset)). The combined intensity of the central beam and Bragg reflections allowed us to estimate crystal thickness at each scan point (Supplementary Figure 7). This estimated thickness correlates well with features observed in annular dark field STEM images including empty regions within holes and the carbon support (Supplementary Figure 7).

*Analysis of lattice structure in 4DSTEM scans across large areas of peptide crystals*
To better understand changes in the pattern of Bragg reflections observed across different areas of a single crystal, we performed unsupervised classification of patterns in each scan (Fig. 2.). Given the high sparsity of signal outside of the central disk in individual diffraction patterns (Supplementary Figure 4, 5), we masked out the central beam to limit its influence on pattern classification (Fig. 2.). This procedure revealed groups of patterns within a scan that shared a common lattice; patterns could be classified in this way for all measured crystals. The number of clusters identified varied between crystals and ranged from 4 to 20 with an average size of $1.6 \times 10^5 \pm 1.7 \times 10^5$ nm$^2$ (Supplementary Figure 8.). In two of the thirty-four datasets analysed, the number of clusters was manually assigned because fewer than 5 clusters were reproducibly identified. The spatial distribution of clusters identified by this approach was consistent with changes in the pattern of Bragg reflections across a scan, rather than other background signal or other diffuse scattering (Supplementary Figure 9.).

The number and intensity of Bragg reflections varied across classified patterns (Fig. 3c, Supplementary Figure 10). Bragg reflections were weak or attenuated in regions where crystals appeared thickest and in thin regions where the central beam showed high counts (Fig. 3c, Supplementary Figure 10). The overall pattern of Bragg reflections differed across a single crystal, changing in a coordinated fashion at all resolutions. Diffraction patterns in a cluster appeared spatially linked within a crystal (Fig. 3b). This effect was reinforced by the requirement that patterns be assigned to a particular class. When mapped onto crystals, diffraction pattern clusters gave the appearance of nanodomains with definite boundaries (Fig. 3a, Supplementary Figure 8b.). However, these boundaries were deceptive since in reality, the change in diffraction appeared more continuous (Supplementary Figure 10.). To assess the true extent of angular reorientation we performed a library-based indexing of patterns obtained by cluster averaging (Supplementary Figure 11). Library indexing of patterns showed that changes in diffraction across clusters could be attributed to a $\pm 3°$ rotation of the lattice away from a common zone (Figure 3c).

*Analysis of lattice structure in HRTEM images of peptide crystals*
To obtain context for the spatial distribution of changes in lattice structure in peptide crystals we observed by 4DSTEM, we performed HRTEM of similar crystals in search of similar effects[20]. High resolution cryo-EM on crystals of QYNNQNNFV with a total dose of 27.8 $e^-/Å^2$ per image (Supplementary Figure 12a,c) allowed us to directly visualize the lattice in these crystals at approximately 2.3Å resolution (Supplementary Figure 12b,d). We processed these images by single particle cryo-EM protocols, identifying image sub-regions for unsupervised clustering, analogous to the procedure used for 4DSTEM pattern classification (Supplementary Figure 13.).

Fourier Transforms of whole crystal images showed clear variation in Bragg reflections when different crystallites are imaged or when comparing crystalline features to empty regions within holes or regions of carbon support (Supplementary Figure 12, 13). Cluster averages also showed differing images (Figure 13c, f, Supplementary Figure 14a - d), but quantifiable differences between clusters were difficult to detect in either real or reciprocal space (Supplementary Figure 14a - h). This is potentially due to the influence of crystal thickness and defocus across the image that make true changes in three-dimensional lattice character difficult to discern (Supplementary Figure 13).

**Discussion**

To reveal the lattice substructure within beam-sensitive three-dimensional crystals, we performed 4DSTEM on peptide nanocrystals at cryogenic temperatures. In contrast to the semi-crystalline polymers studied previously by 4DSTEM[41,42], the prion peptide nanocrystals we evaluated are composed of highly ordered peptide arrays that diffract to sub-ångstrom resolution[33]. Taking advantage of the known lattice parameters for these crystals, their highly ordered structure, and the known atomic arrangement of their constituent molecules, we probed for nano-scale changes in diffraction across single crystals. The data we obtained add to previous studies on the lattice substructure and physical properties of nanometre to micrometre sized protein crystals investigated by a variety of methods[6,20,25,26,57].

Unlike other methods of protein crystal characterization[20,58–60], 4DSTEM provides direct observation of lattice character through nanoscale mapping of changes in diffraction across micron scale areas[41]. Our measurement of 1.4Å resolution diffraction from sub-10nm regions of peptide crystals was facilitated by three key technological features of our experiment: (i) fast readout direct electron counting detectors[61], (ii) a hybrid counting protocol applied to sparse diffraction data captured by 4DSTEM, and (iii) low dose cryogenic techniques that lessen the evidence of radiation damage. Because of these key features, the resolution we achieve in 4DSTEM scans is comparable to that of diffraction patterns measured by MicroED using a selected area aperture. The fluence required to achieve this resolution by 4DSTEM is higher than it is for MicroED, on the order of 1 e$^-$/Å$^2$ for the former compared to about 0.01 e$^-$/Å$^2$ for the latter [62]. The dose chosen for these 4DSTEM experiments is necessary to achieve high resolution diffraction from a significantly lower number of molecules diffracted at each scan point (~1-10x10$^3$ molecules) compared to those diffracted by MicroED from similar crystals (~1x10$^7$ molecules), even for the smallest selected area apertures [27,35]. The higher potential for damage by this dose at each scan position is mitigated by spacing scan steps by a distance larger than the probe size, ensuring an unprobed region of the crystal is illuminated with each scan step. While several step sizes were explored across scans, a step size of 20nm was sufficiently fine to allow pattern classification yet large enough to avoid perceptible pattern degradation due to damage from neighbouring regions.

The changes in diffraction we observe across micron-sized regions of peptide nanocrystals point to inhomogeneities across a single crystal. These differences are not only diagnostic of regions with strong and weak diffraction within individual crystals, but also point to changes in the orientation of the lattice within a crystal. These lattice nano-ripples in sub-micron regions of a single nanocrystal are difficult to detect by methods that make use of selected area electron diffraction, as well as those that integrate diffraction from whole crystals. This is especially true after data from multiple crystals is merged[63,64], a requirement for both serial crystallography and most MicroED experiments[35]. Lattice bending at this scale in macromolecular nanocrystals may explain the discrepancy in dynamical scattering observed from crystals of this thickness by MicroED, compared to what would be expected by simulation from perfect crystals[65].

The lattice nano-ripples observed by 4DSTEM differ from conventional domain blocks. Whereas conventional mosaic blocks have discrete boundaries and a semi-random distribution of orientations, we observe a more gradual, progressive change in orientation as a function of spatial localization. Presently, our library-based orientation assignment acts on the average diffraction over clustered patterns and requires known cell constants, but improved interpretation of single sparse patterns and *ab initio* indexing of electron diffraction patterns may alleviate one or both of these limitations. Despite present limitations, nano-scale lattice changes in our crystals are more readily identifiable from 4DSTEM scans than from HRTEM images of similar nanocrystals owing to the higher contrast of Bragg peaks captured by diffraction. Transforms of sub-regions within a 3D lysozyme nanocrystal have shown similar lattice differences across a single crystal[20]. However, in that case, the changes were also potentially attributable to differences in defocus across the crystal or to changes in the level of dynamical scattering across non uniformly thick areas of the crystal[20]. Interpretation of our own images of frozen-hydrated peptide nanocrystals presents similar challenges (Supplementary Figure 12, 13). 4DSTEM allows a quantitative classification of lattice orientations within a crystal and can reveal more subtle changes than those identified in HRTEM images of similar crystals. This difference may point to a greater sensitivity to detection of lattice changes by 4DSTEM at even lower doses than those required for HRTEM.

Lastly, by allowing nanoscale (20nm) interrogation of lattice structure in macromolecular crystals, 4DSTEM offers a potential way to shrink the number of molecules required for structure determination by electron diffraction from single or serial crystal data.

**Methods**

*Preparation of crystals*
Crystals were grown as previously described[33]. Briefly, synthetic peptide (GenScript) with sequence QYNNQNNFV at a purity of 98% was dissolved in ultrapure water to a final concentration of 3.5mM. QYNNQNNFV crystals were formed in batch by mixing dissolved peptide solution 1:1 with a buffer solution composed of 10% MPD and 0.1M MES at pH 6.0.

*4DSTEM Data collection strategy*

Crystal suspensions were diluted two-fold in water or crystallization buffer before being dispensed onto holey carbon grids (Quantifoil 2/4, #300 copper; Ted Pella Inc) and allowed to air dry. Samples were introduced into the TEAM I microscope (FEI Titan) with a Gatan 636 Cryo holder, cooled to liquid nitrogen temperatures throughout all data acquisition. The TEAM I microscope was operated at 300 keV in STEM mode with a convergence half-angle of 0.5 mrad resulting in a ~6nm convergent beam (Supplementary Figure 2).

Crystals of interest were identified using annular dark-field STEM at low magnification (Fig 1a (inset).). For this, the electron fluence (dose, $e^-/Å^2$) was limited to a fluence of less than 0.01 $e^-/Å^2$ to minimize diffraction decay due to radiation damage, starting with parameters identified for other beam sensitive organic lattices[41] (Supplementary Table 1.). The search for crystals was performed by scanning a focussed beam with a dwell time of 1 microsecond over a 512 x 512 scans across 14 x 14 μm fields of view, corresponding to an electron fluence <0.01 $e^-/A^2$. We chose single crystals or crystal bundles visible at low magnification in annular dark field STEM images for further evaluation by 4DSTEM (Fig 1a-c.).

4DSTEM datasets were collected by scanning the 6nm probe over the sample in two-dimensional scan using a 2.5 ms/step dwell time and a 20nm step size. The electrostatic lens above the probe forming aperture was adjusted such that the overall fluence per scan step was ~1 $e^-/Å^2$. Diffraction patterns were recorded using a Gatan K2-IS direct electron detector (GATAN), with a 1792 x 1920 pixel detection and an effective camera length of 575 mm (Fig. 1a.). Typical scans consisted of 1,000 to 10,000 diffraction patterns captured at 400 frames/second; datasets were ~ 5 - 30GB in size.

*4DSTEM Dose estimation*

The total dose imparted per 4DSTEM scan step was estimated as follows: The convergent probe was imaged in TEM mode at high magnification using a Gatan US1000 CCD detector with an exposure time of 10 seconds and all other beam settings the same as those used for data acquisition (Supplementary Figure 2). Using a nominal conversion rate of 3 counts per $e^-$ at 300 keV, we estimated the total number of electrons within the probe to be $1.5 \times 10^6$ $e^-/s$. This number calculated from the image of the probe agrees with the screen current estimate for a similar gun lens setting. The diameter of the probe at FWHM was measured to be 6nm (Supplementary Figure 2) and, given an exposure time of 0.0025s, the final dose per scan step was estimated as ~1$e^-/Å^2$. The calculated dose above is different than that calculated from the field of view which would have been reported as an order of magnitude less due to the real space 'pixel size' of 15-20nm.

*Image processing and hybrid counting of 4DSTEM data*

All data processing was performed using custom scripts written in the MATLAB (MathWorks) programming language. Patterns belonging to a single area scan were jointly processed to achieve hybrid counting as follows. We computed an average pattern from all of the images within each dataset (Supplementary Figure 4). We estimated a differential dark current offset between detector strips as the median value of each strip of pixels in the vertical direction of the measured patterns (Supplementary Figure 4c). These values were subsequently subtracted from each diffraction pattern within the dataset (Supplementary Figure 4d). After dark current correction, a Gaussian background was fit to the distribution of background subtracted pixel values of all patterns within the dataset (Supplementary Figure 4e). This fit was used to estimate a threshold for the detector dark noise such that counts that were below 5 standard deviations of the Gaussian fit were considered background (Supplementary Figure 4e). This threshold would typically be used in standard electron counting algorithms[16,17,66].

To recover some of the dynamic range lost due to coincident electron events we implemented a 'hybrid counting' approach. Here we divided all pixel values by the calculated threshold and the floor function of these values was taken to give an estimate of the degree of coincident electron events occurring at individual pixels on the detector. (Supplementary Figure 5). We found that correcting in this way resulted in the majority of events being either single or double counts and reduced the noise floor to close to zero, crucial for accurate clustering. However, in some cases where there was direct transmission of the electron beam, where the scan passed over vacuum, counts were much higher (Supplementary Figure 5 (inset)). Whilst the assumption of a linear relationship between the detector counts and electron coincidence is incorrect, this method better captured the true transmission of the central beam than considering all counts above the threshold as single electron events. This was important for more accurately estimating crystal thickness. Finally, zero values were discarded and the 2D images were converted to coordinate lists and corresponding electron counts; a ~600-fold reduction in data size.

*Correction of diffraction shift in 4DSTEM scans*

The horizontal and vertical shift of diffraction patterns induced by beam tilt during these relatively low magnification scans was corrected by tracking the centre of mass of the transmitted beam. As the signal to noise in a single pattern was not sufficient to calculate this accurately we separated the problem into two steps, independently correcting shifts in the x-scan and y-scan directions. A strip wise average of patterns was taken along the direction to be corrected; for example, for a 72x50 scan we would first average patterns along the first dimension to give 1x50 strip wise averaged patterns. The centre of mass of the transmitted beam was used to give an estimate of the pixel shift for each strip and individual datasets within this strip were shifted to a common centre based on this. This process was repeated in the second dimension. For most datasets, just one round of shift correction was sufficient but in cases where the shift was particularly problematic several rounds were necessary (Supplementary Figure 6).

*Estimation of crystal thickness*

Crystal thickness was estimated using the log-ratio formula typically employed in EELS for inelastic scattering:

$$Z_{xy} = -\lambda \cdot \ln\left(\frac{I_{xy}}{I_0}\right)$$

Where $Z_{xy}$ is the thickness of the crystal at scan location *xy* and $\lambda$ is the mean free path of electrons through the peptide crystals. A lambda value of 332nm was used based on estimates previously determined from equivalent crystals[67]. $I_{xy}$ is the integrated transmitted beam at scan position *xy* (Supplementary Figure 7a.) plus the integrated intensity at all Bragg peak positions (Supplementary Figure 7b.). This sum represents intensity variation due to changes in inelastic scattering and was used for thickness estimates (Supplementary Figure 7g.). $I_0$ is calculated as the mean of the integrated central beam at scan positions that were over vacuum minus two times the standard deviation of values in this region. This correction was to account for fluctuations observed in the central beam intensity.

*HRTEM data collection and image processing*

A monodisperse solution of QYNNQNNFV crystals was applied onto holey carbon grids (Quantifoil 1/4, #300 copper; Ted Pella Inc) and plunge-frozen using a Vitrobot Mark IV robot (ThermoFisher Scientific). Data was collected on a CS aberration corrected FEI Titan Krios (ThermoFisher Scientific) operated at 300 KeV. Super-resolution movies were recorded using a Gatan K2 Summit direct electron detector. The nominal physical pixel size was 1.04 Å/pixel (0.52 Å/pixel in super-resolution movie frames) and dose per frame was 1.39 e$^-$/Å$^2$. A total of 20 frames were taken for each movie resulting in a final dose of 27.8 e$^-$/Å$^2$ per image. Super-resolution movie stacks were corrected for gain and beam-induced motion using MotionCorr2[68]. Initial estimates of anisotropic magnification were made using 10 micrographs where crystalline ice was present using the program mag-distortion-estimate[69]. These parameters were used to correct for anisotropic magnification in MotionCorr2. Frames were subsequently aligned without using patches, dose-weighted, and down-sampled by two.

*Lattice mapping in 4DSTEM and HRTEM*

Before lattice mapping individual diffraction patterns were binned 8-fold to increase their SNR. The central beam was then masked out to remove the influence of transmission from the lattice mapping; we noticed that without this step the lattice maps produced matched variations in the thickness of the crystal too closely. We used *k-means* clustering to sort diffraction patterns from a single 4DSTEM dataset into different clusters based on their Euclidian distance from the average pattern within a particular cluster:

$$\underset{S}{\operatorname{argmin}} \sum_{i=1}^{K} \sum_{x \in S_i} \|x - \mu i\|^2$$

Where *K* is the number of clusters determined using an implementation of the *G-means* algorithm with a = 0.001[70]. $S_i$ is an individual cluster, $\mu_I$ is the current average of all patterns within the cluster $S_i$ and *x* is an individual diffraction pattern within the 4DSTEM dataset. The *k-means* ++ algorithm was used to initialize $\mu_i$ for each cluster[71]. The algorithm was stopped when either 100 iterations were performed or when the within-cluster sum of squares (WCSS) score stagnated. Clusters where the beam passed over vacuum showed little signal when the primary beam was excluded, preventing distances between patterns in this cluster from showing Gaussian behaviour; their Euclidean distances from the mean would be close to unity. To circumvent this, we included a break in our implementation of *G-means* that stopped assignment of clusters once at least 80% of the assigned clusters were found to be Gaussian by the Anderson-Darling statistic. This substantially improved convergence of the algorithm.

For real space images collected by HRTEM a similar procedure was followed with the following modifications. First 128 by 128 sub-images were cropped from the larger HRTEM images with no overlapping regions. We estimated K by the elbow method[72], which we found to be more stable than the Anderson-Darling statistic for the noisier imaging data. The clustering step was applied as above, with the inclusion of a 5 by 5 pixel-wise search to account for potential misalignments between the two images under consideration.

*Indexing of lattices in 4DSTEM cluster averages*

To assess the underlying lattice reorganization that could account for the changes in diffraction observed from the clustering, we indexed the cluster averages via library matching (Supplementary figure 11). A library of NBED patterns were simulated with PRISM[73] using the known crystal structure of QYNNQNFV. We simulated expected NBED patterns arising from a 40 x 40nm region of a perfect crystal of varying thickness (10 – 600 nm). Probe size and convergence semi-angle were fixed to match experimental parameters and patterns were calculated at a range of *xy* tilts, ± 4 degrees in 0.25-degree increments, away from the *hk0* zone or mean orientation in cases where the crystal was not sitting close to a zone axis. To match the experimental patterns to the simulated library, the positions of all possible peaks that could arise within the bounds of the HAADF detector were identified, excluding the central disk. These peak locations were then used to create a list of intensities by integrating all pixel values within a 4-pixel radius centred around each peaks *kxky* position. For each cluster average and simulated pattern, the listed intensities were scaled to be a ratio of the most intense peak within that pattern. Intensity lists were then compared by RMSD of intensity values at all peak positions within a given pair of patterns such that:

$$RMSD_{ij} = \sqrt{\frac{\sum_{p=1}^{P}(\mu i_p - Sim j_p)^2}{P}}$$

Where *p* represents an individual peak position from the set of peak positions *P*, *μi* is the *ith* averaged cluster pattern and *Simj* is the *jth* pattern within the simulated library. The orientation of the simulated pattern with the lowest RMSD to the current cluster average was then assigned to the orientation of the lattice within that region.

*Fourier filtering of HRTEM images*

To improve the contrast of the average lattice images captured by HRTEM Fourier filtering was performed in a similar manner to Erickson and Klug[74]. For each class average the Fourier transform was computed and the mean and standard deviation were calculated for the magnitude of all pixel values, excluding the DC term. All pixels in the Fourier transform with a magnitude lower than 2 standard deviations above the mean magnitude were set to zero and the final filtered image was calculated by inverse Fourier transform.

**Data Avaliability**

The MATLAB scripts for data pre-processing, clustering, orientation assignment as well as some of the pre-processed data used for clustering and orientation assignment can be found at: https://github.com/marcusgj13/4DSTEM_dataAnalysis. The raw .dm4 files will be available at https://www.ebi.ac.uk/pdbe/emdb/empiar/

**Acknowledgments:** We thank Zhiheng Yu and Rick Huang (Janelia Research Campus) for their technical assistance in HR-TEM data collection; Peter Ercius (NCEM); David Eisenberg, Tamir Gonen, Johan Hattne, Michael Martynowycz, Michael Sawaya, Duilio Cascio (UCLA) and Wolfgang Kabsch (MPI) for insightful discussions. This work was supported in part by STROBE: A National Science Foundation Science and Technology

Center under Grant No. DMR 1548924 and DOE Grant DE-FC02-02ER63421. 4DSTEM was performed as a user project at the Molecular Foundry at Lawrence Berkeley National Laboratory, which is supported by the U.S. Department of Energy under Contract # DE-AC02-05CH11231. M.G.J is supported by a QCB Collaboratory Postdoctoral Fellowship. J.A.R. is supported as a Beckman Young Investigator, a Searle Scholar and a Pew Scholar. O.P. is supported by the Electron Microscopy of Soft Matter Program from the Office of Science, Office of Basic Energy Sciences, Materials Sciences and Engineering Division of the U.S. Department of Energy under Contract No. DE-AC02-05CH11231. J.C. acknowledges support from the U.S. Department of Energy Early Career Research Program. C.G. is supported by a Ruth L. Kirschstein National Research Service Award GM007185. D.R.B is supported by a National Science Foundation Graduate Research Fellowship.

**Author contributions:** J.A.R. and A.M. conceived of the project. J.A.R. directed the work. C.G., M.G.J. and J.A.R. grew, evaluated and optimized crystals. M.G.J., D.R.B., C.G., K.C.B., O.P., J.C., C.Z. and J.A.R. collected data. M.G.J., C.O., K.C.B., K.C.M. D.R.B. and J.A.R. analyzed the data. M.G.J. and J.A.R. wrote the article, with input from all authors.

**Correspondance :** Jose A. Rodriguez; email: jrodriguez@mbi.ucla.edu

**Competing interests:** The authors declare no competing interests.

**Figures**

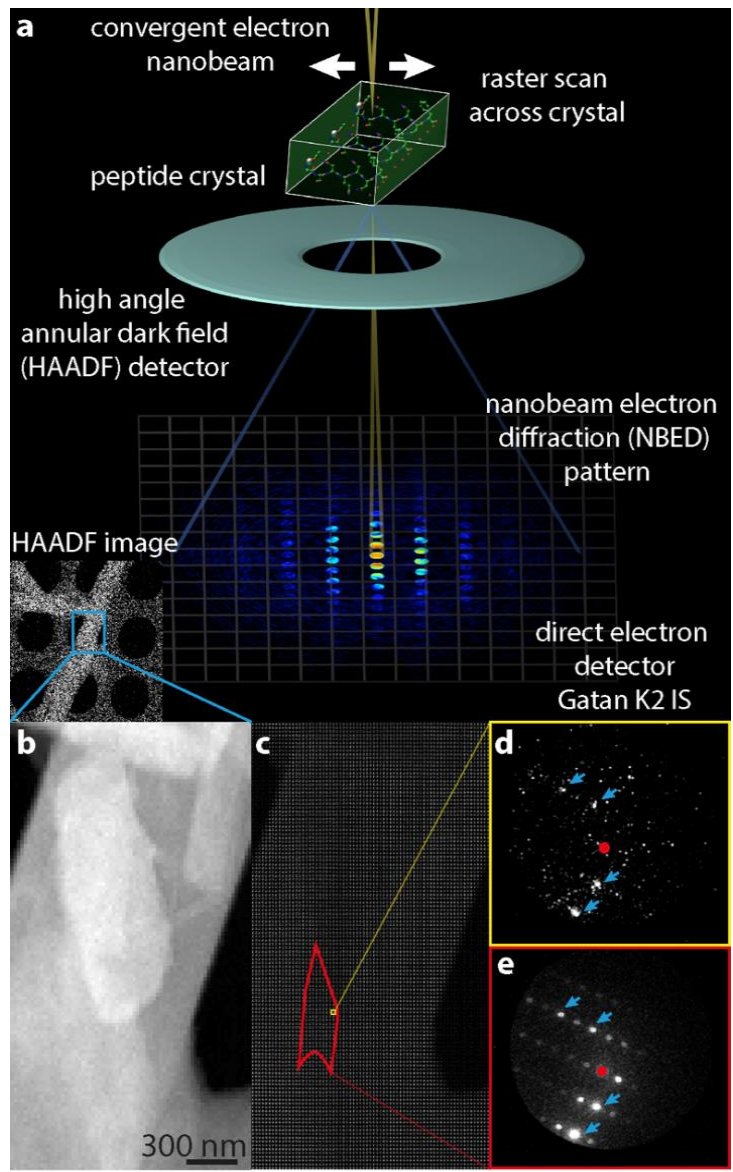

Fig. 1. **Measuring lattice structure in peptide nanocrystals by 4DSTEM**. (**a**) Diagram of 4DSTEM experiment shows key aspects and components of the procedure; inset shows a low dose and low magnification high angle annular dark field (HAADF) STEM image. (**b**) A higher resolution STEM image shows the crystal in inset of (**a**) with greater detail. (**c**) Image montage shows all patterns collected in a 4DSTEM scan captured alongside the image in (**b**); a single diffraction pattern is shown in (**d**), and an average of all diffraction patterns from the red region highlighted in (**c**) is shown in (**e**). Primary beam is masked by red circles, blue arrows indicate a subset of Bragg reflections.

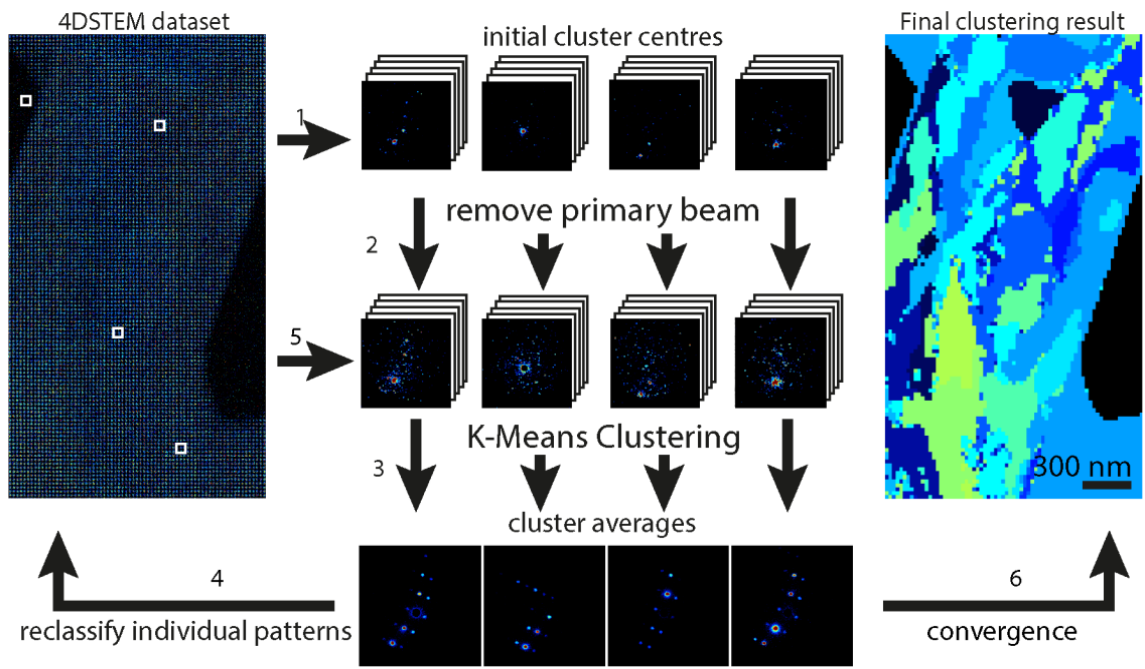

Fig 2. **Workflow of unsupervised clustering used to define nanoripples in peptide crystals.** (step 1) Initial cluster "centres" are assigned within the 4DSTEM dataset (left white boxes) and all patterns are binned. (step 2) The primary beam is subsequently masked to prevent its influence on clustering. (step 3) Individual patterns are then compared to each cluster centre sequentially via Euclidean distance and assigned to the cluster where this distance is smallest. (step 4) Average patterns are calculated for each cluster and (step 5) are used to reclassify individual patterns until convergence (step 6) where a final map illustrating spatial localization of similar diffraction patterns is produced.

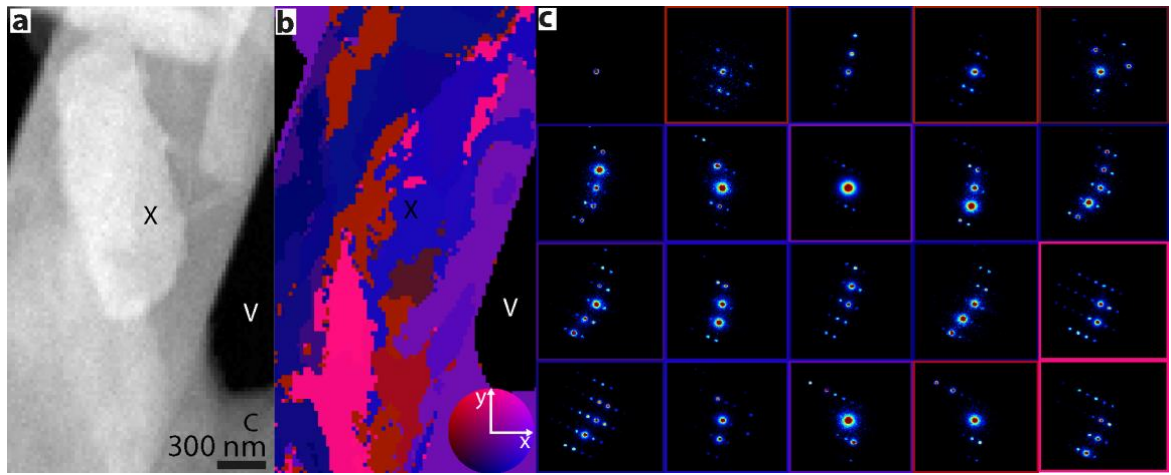

Fig 3. **Mapping of lattice nano-ripples within peptide nanocrystals by 4DSTEM.** (a) HAADF image of a QYNNQNNFV nanocrystal. Unsupervised classification of diffraction patterns captured by the 4DSTEM scan acquired during the measurement of (a) are shown in (b); this is a map of diffraction clusters not obvious from (a). Colours in (b) illustrate the change in lattice orientation for each individual cluster with respect to the mean orientation across the scan area. The colour wheel (inset) demonstrates the relative orientation away from the mean in x and y-tilt, with a maximum deviation of 4 degrees. Average diffraction patterns from diffraction outlined in (b) are shown in (c), where the colour of the bounding box corresponds to the colour of the corresponding cluster in (b). C, X and V indicate carbon support, peptide crystal and vacuum respectively.

## Supplementary Tables

Supplementary Table 1. **Experimental parameters for measurement of electron diffraction of ordered molecular assemblies**

|  | Panova et al (2016) | This experiment | MicroED |
|---|---|---|---|
| Instrument | FEI Titan | TEAM I | FEI Tecnai F20 |
| Detector | Gatan Orius | Gatan K2 IS | TVIPS F416 |
| Temperature (K) | 77 | 77 | 77 |
| Accelerating voltage (keV) | 200 | 300 | 200 |
| Flux (e$^-$/s) | 3.2x10$^7$ | 1.5x10$^6$ | 2.4x10$^5$ |
| Probe size (nm at FWHM) | 7 | 6 | 1000 (SA aperture) |
| Probe convergence angle (mrad) | 0.51 | 0.5 | near parallel |
| Estimated dose (e$^-$/Å$^2$) | 600 | 1 | 0.01 |
| Dwell time (ms) | 70 | 2.5 | 3000 |
| Step size (nm) | 20 – 40 | 15 - 20 | N/A |

**Supplementary Figures**

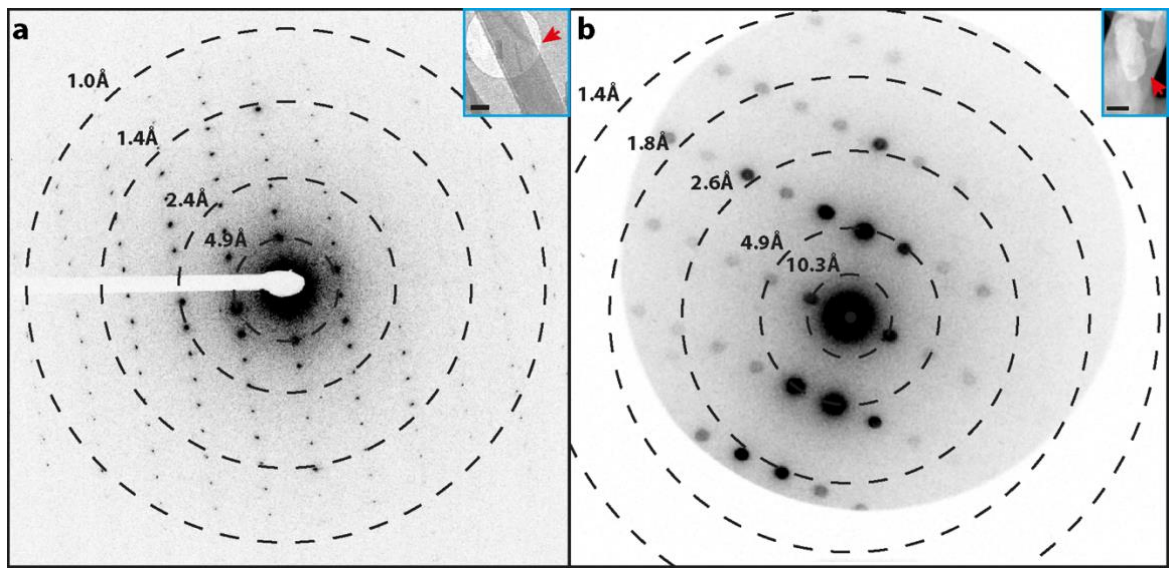

Supplementary Figure 1. **Comparison of diffraction patterns collected by MicroED and 4DSTEM.** (**a**) A diffraction photograph measured by continuous rotation MicroED from a frozen hydrated crystal of QYNNQNNFV (inset). (**b**) Diffraction pattern calculated by taking the average of all patterns within a single 4DSTEM dataset acquired from a QYNNQNNFV crystal (inset) from the same batch as (**a**). Scale bars are 300 nm in all images.

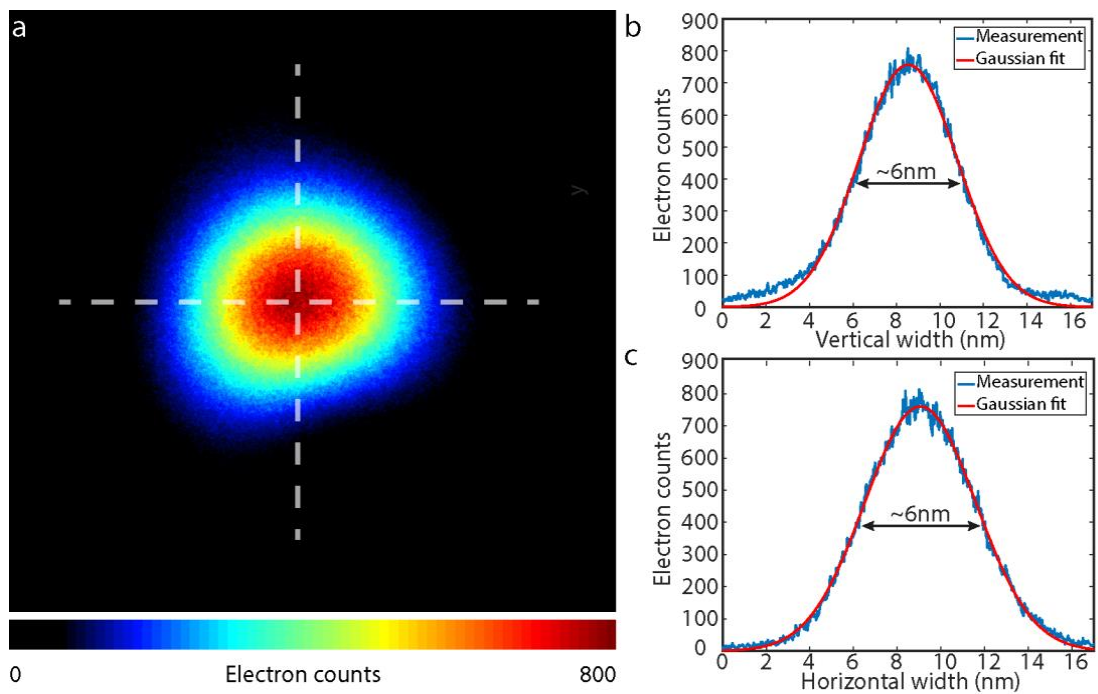

Supplementary Figure 2. **Measure of probe properties used for 4DSTEM.** (a) Image of the probe captured on a Gatan Ultra scan 1000 with an exposure time of 10s. (b) Vertical line scan across the probe (white dashed line in (a)) and Gaussian fit to the profile giving a diameter of ~6nm at FWHM. (c) Horizontal line scan across the probe (white dashed line in (a)) and Gaussian fit to the profile giving a diameter of ~6nm at FWHM.

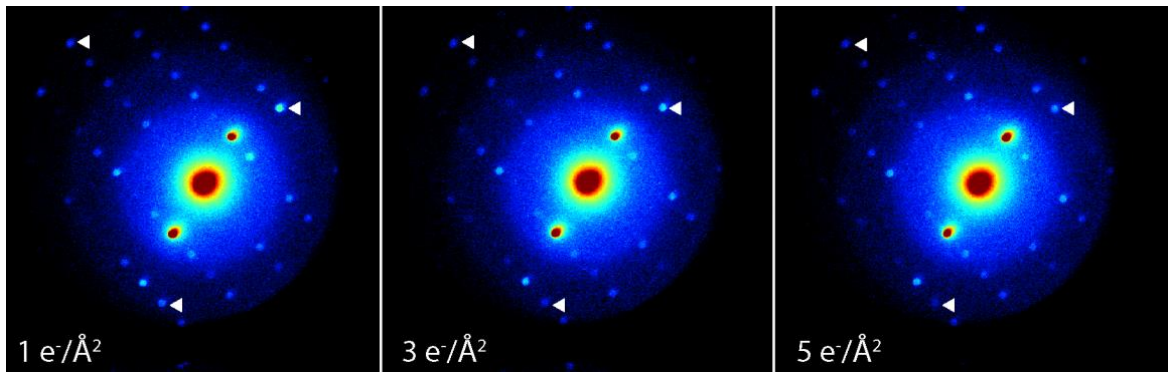

Supplementary Figure 3. **Changes in Bragg reflection brightness as a function of electron dose in 4DSTEM experiments.** (left to right) Average diffraction patterns of 4DSTEM datasets successively acquired from the same crystal (left: first pass, middle: third pass, right: fifth pass) with the total accumulated dose below. White arrow indicates high resolution Bragg reflections whose intensity decreases as a function of dose.

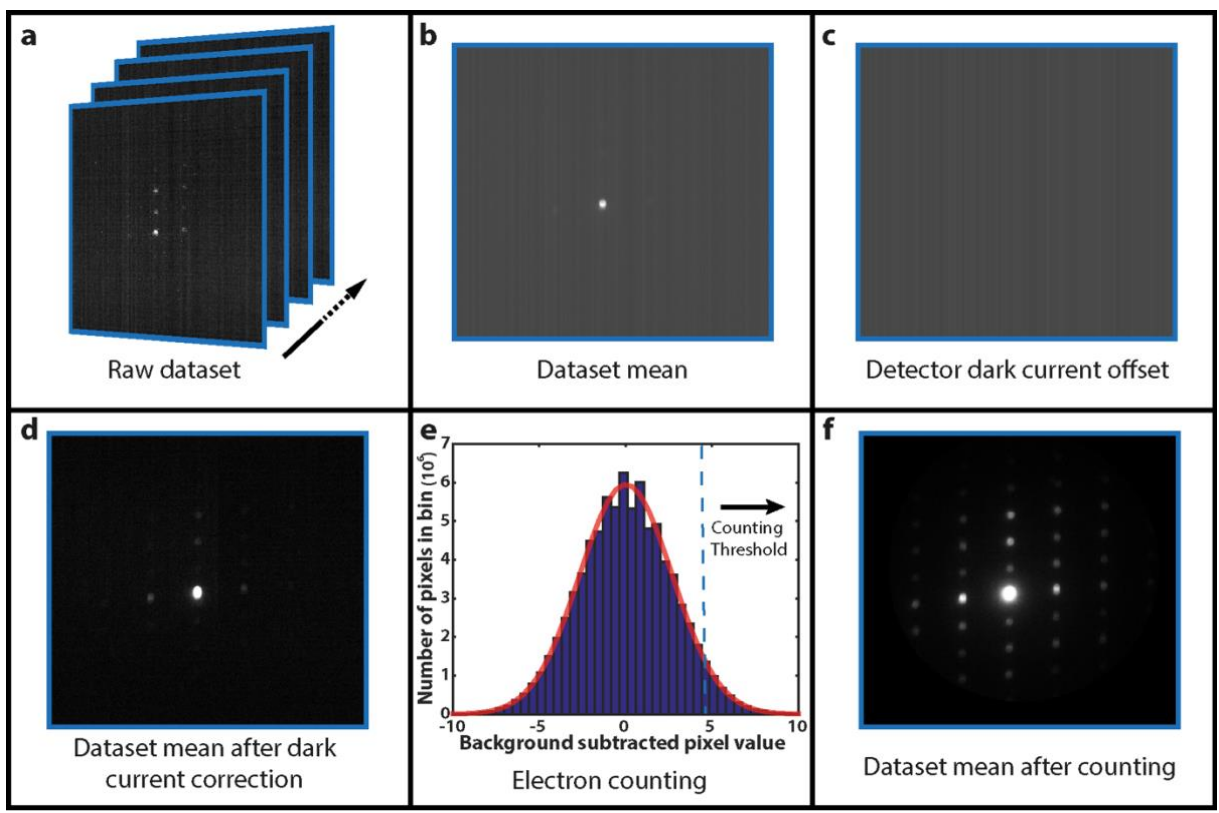

Supplementary Figure 4. **Data processing pipeline for 4DSTEM data.** (a) Raw diffraction patterns recorded at 400 frames per second are first combined to obtain a mean pattern (b). A dark current offset across the detector is estimated from the mean pattern using a median filter (c) and subsequently subtracted from all images (d). A Gaussian background is then fit to the distribution of all pixel values within the background corrected data and based on this a threshold is determined above which the incidence of electrons is considered true (e); intensity values are then converted to electron counts (f).

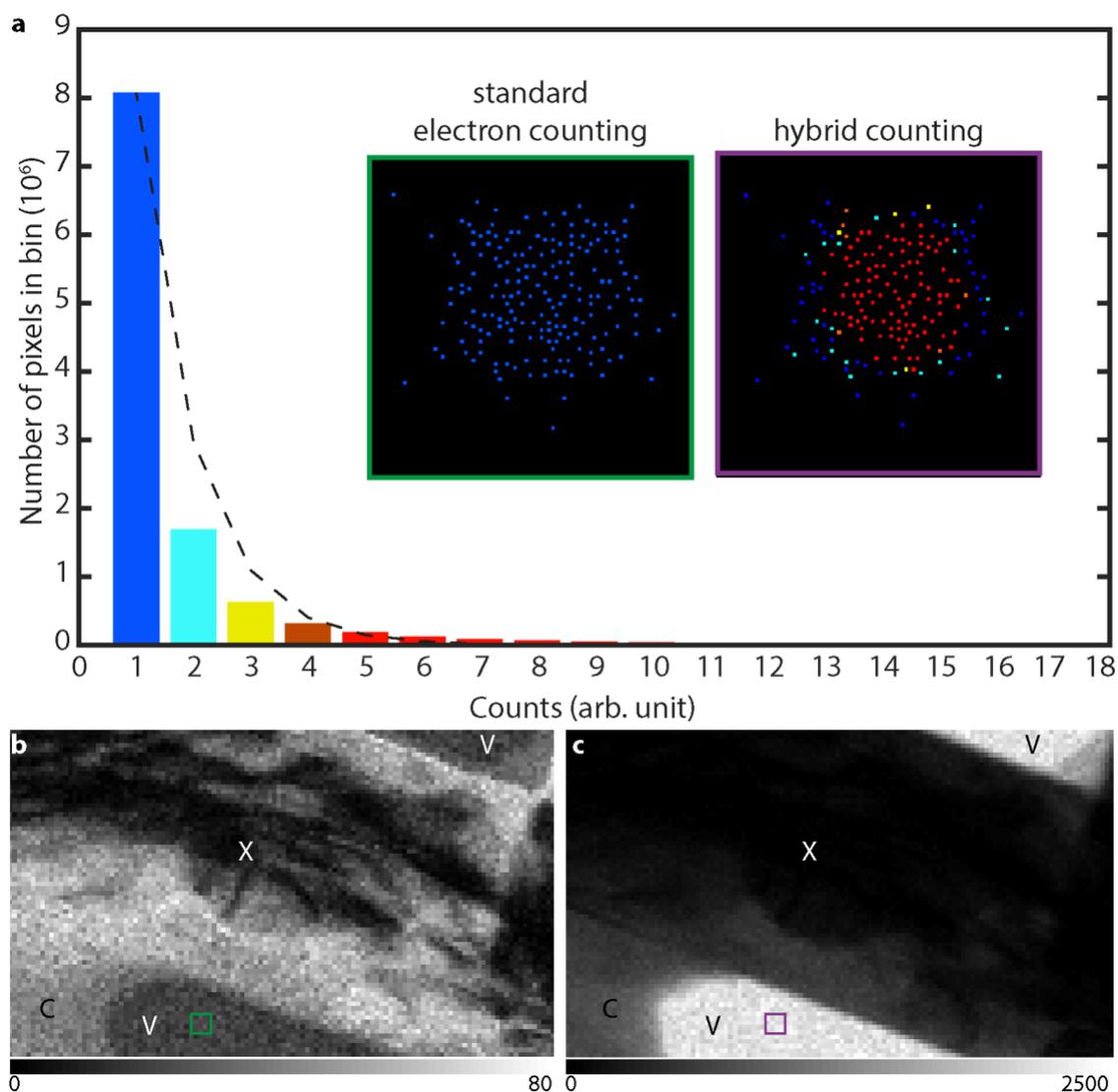

Supplementary Figure 5. **Use of hybrid pixel counts to improve estimates of beam transmission.** (**a**) Histogram of the distribution of counts due to individual and multiple electron events recorded in a representative 4DSTEM dataset after hybrid counting. (inset) Image of the primary beam over vacuum using standard binary electron counting (left) and hybrid counting (right) from the green and purple square region in (**b**) and (**c**) respectively. Colours in insets match the colours in the histogram (images are truncated at 5 counts to improve visibility of low counts). The dashed line shows the best fit of the count histogram to a Poisson distribution. (**b**) Map of central beam intensity using standard electron counting. (**c**) Map of central beam intensity using hybrid counting. C, X and V represent areas over carbon support, peptide crystal and vacuum respectively. Scale bars represent integrated intensity at each pixel position.

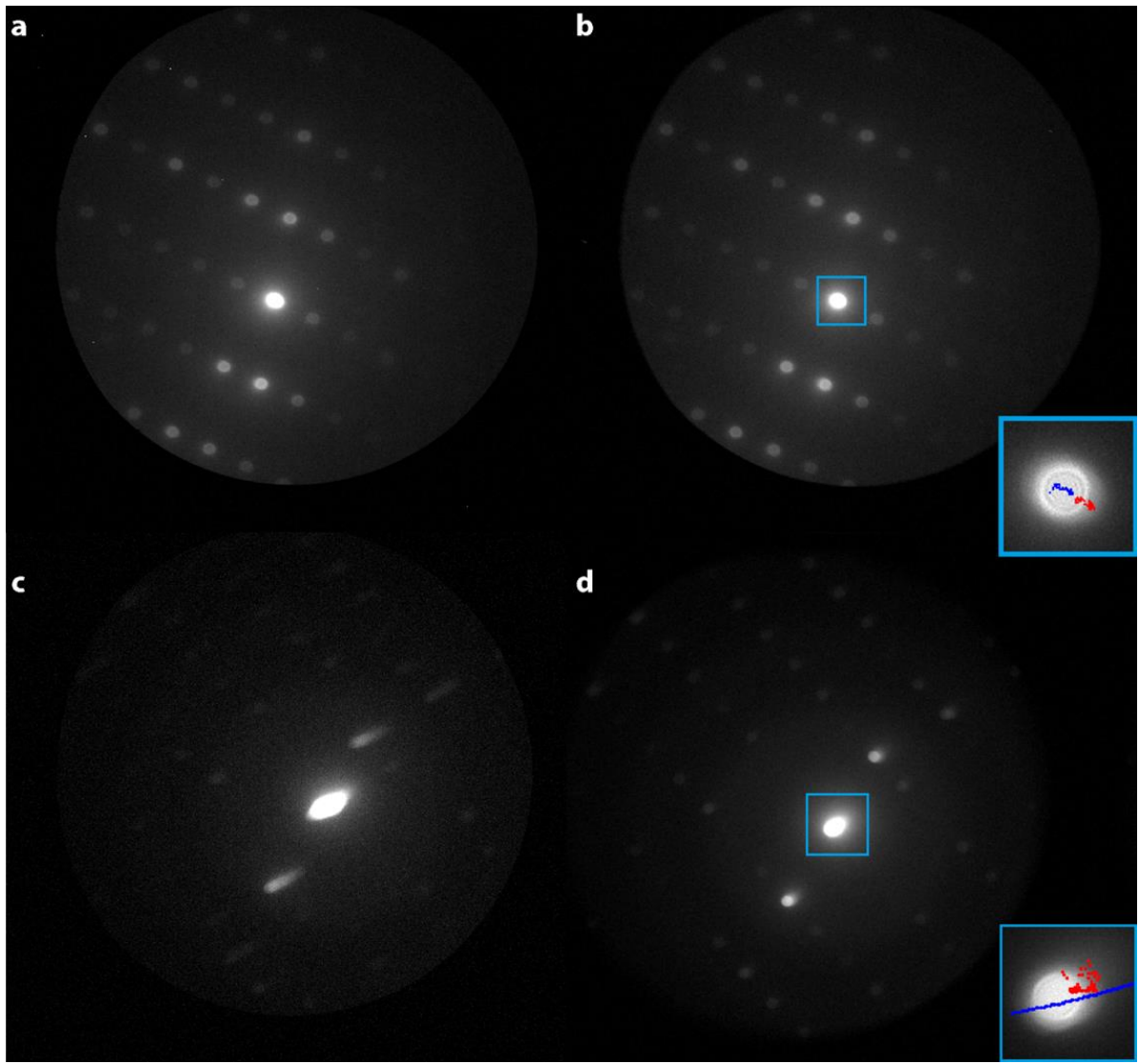

Supplementary Figure 6. **Beam shift correction in 4DSTEM datasets.** Mean diffraction of a 4DSTEM datasets showing minimal beam shift due to tilting of the probe during low magnification scanning acquisition both before (**a**) and after (**b**) shift correction. The amount of drift in first and second scan directions is represented as blue and red circles (inset). (**c**) Dataset showing large beam shift before correction. (**d**) Reduction in beam shift visible after correction. The amount of drift in first and second scan directions is represented as blue and red circles (inset).

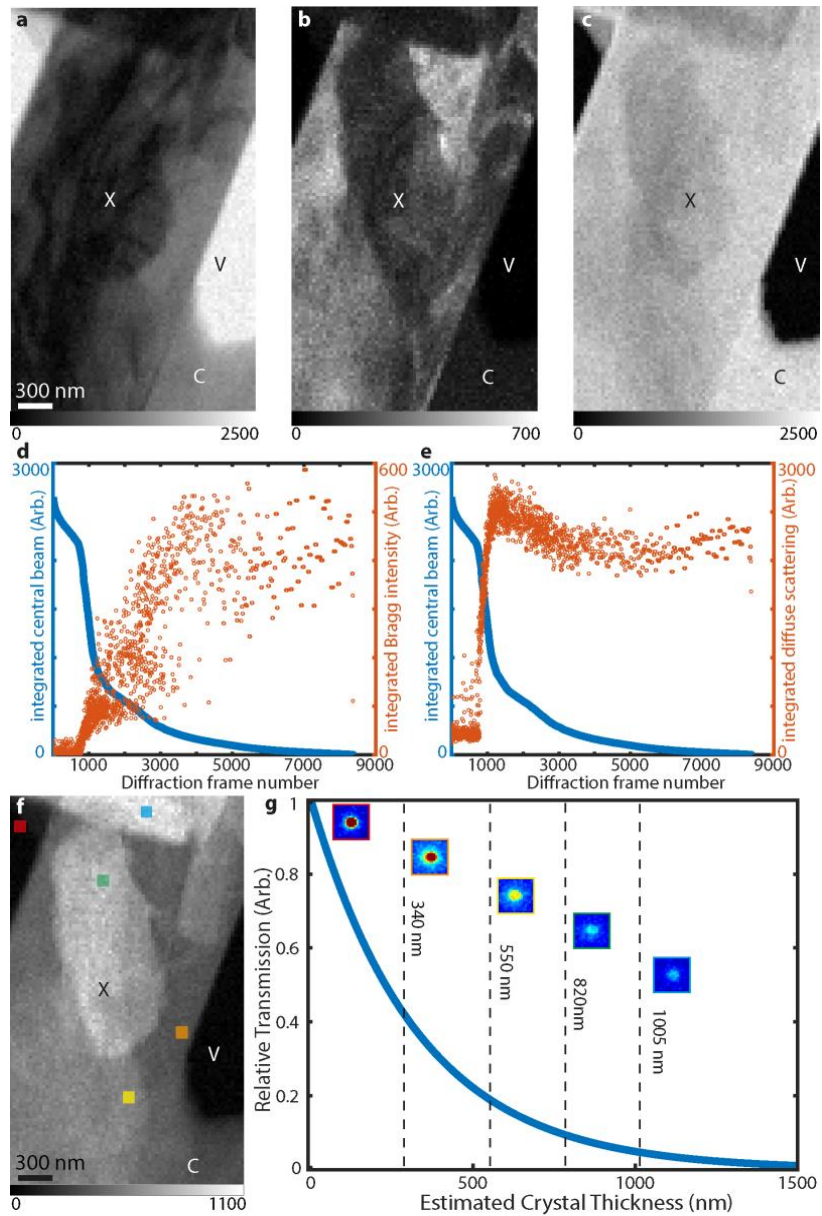

Supplementary Figure 7. **Estimation of crystal thickness from 4DSTEM patterns.** (**a - c**) Images reconstructed by integrating (**a**) the central beam, (**b**) all Bragg reflections or (**c**) the diffuse scattering within each 4DSTEM pattern. Colourbars represent integrated intensity. (**d**) Plot of integrated Bragg peaks as a function of central beam intensity. (**e**) Plot of diffuse background intensity as a function of central beam intensity. (**f**) Reconstructed image of crystal thickness estimated by combining information in (**a**) and (**b**), colourbar represents thickness in nm. (**g**) Plot of focused, central beam transmission as a function of sample thickness in a 4DSTEM scan. Inset images show the primary beam at the positions indicated by correspondingly coloured squares on the thickness map (**f**). All images are on the same intensity scale where red indicates high transmission and blue, low. C, X and V represent areas over carbon support, peptide crystal and vacuum respectively.

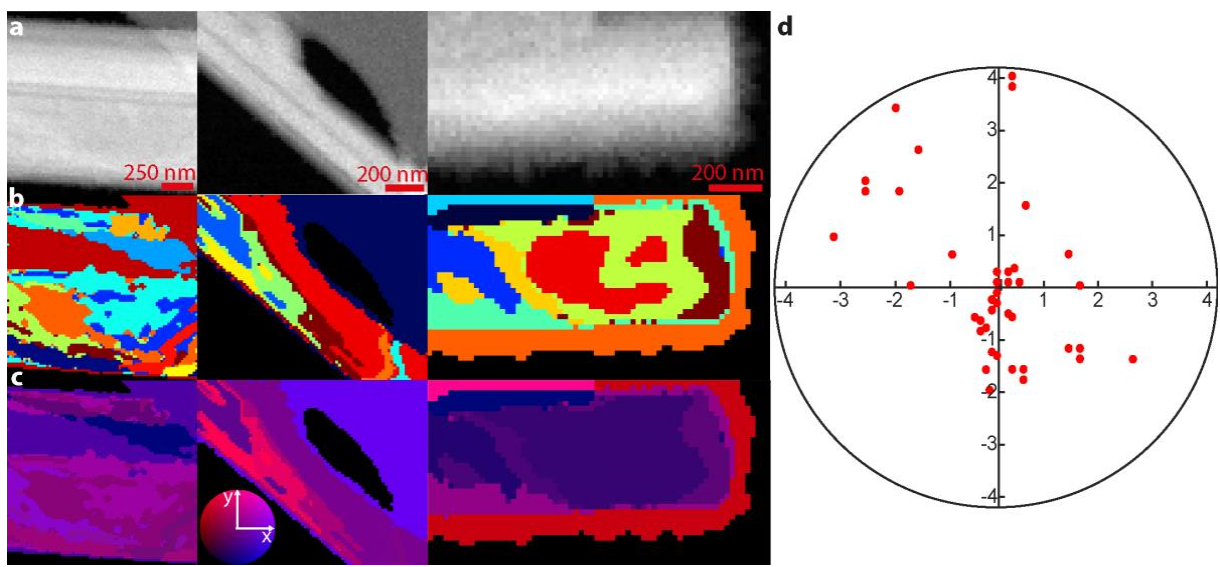

Supplementary Figure 8. **Representative HAADF images and orientation maps acquired from peptide nanocrystals by 4DSTEM.** (**a**) HAADF images of QYNNQNNFV crystals. (**b**) Maps of diffraction localization reconstructed from 4DSTEM scans of the same crystals as (**a**). (**c**) Maps recoloured after orientation assignment. The colour wheel (inset) demonstrates the relative orientation away from the mean in x and y-tilt, with a maximum deviation of 4 degrees. (**d**) Scatter plot of angular deviation from mean crystal orientation identified across QYNNQNNFV crystals analysed by 4DSTEM. Axes represent angular offset in degrees for the x and y-tilt directions respectively.

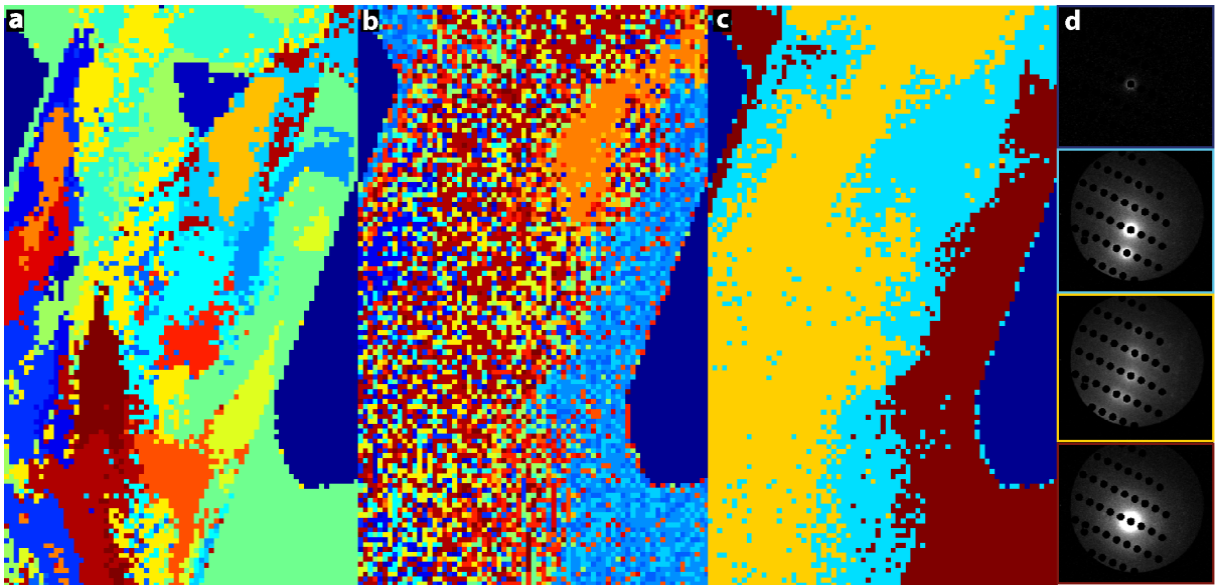

Supplementary Figure 9. **Influence of signal outside Bragg reflections on pattern classification and diffraction mapping.** (**a**) Map of diffraction clusters obtained from the nanocrystal seen in Fig. 2 in which 20 clusters were identified using the complete diffraction pattern (disregarding the primary beam). (**b**) Cluster map with 20 clusters calculated by excluding Bragg reflections - using only the diffuse scattering between Bragg peaks. (**c**) Cluster map calculated as in (**b**) but limited to 4 clusters. When the number of clusters is restricted, the map more closely follows the thickness map of the crystal. (**d**) Cluster averages for each of the 4 clusters shown in (**c**) with Bragg peaks removed.

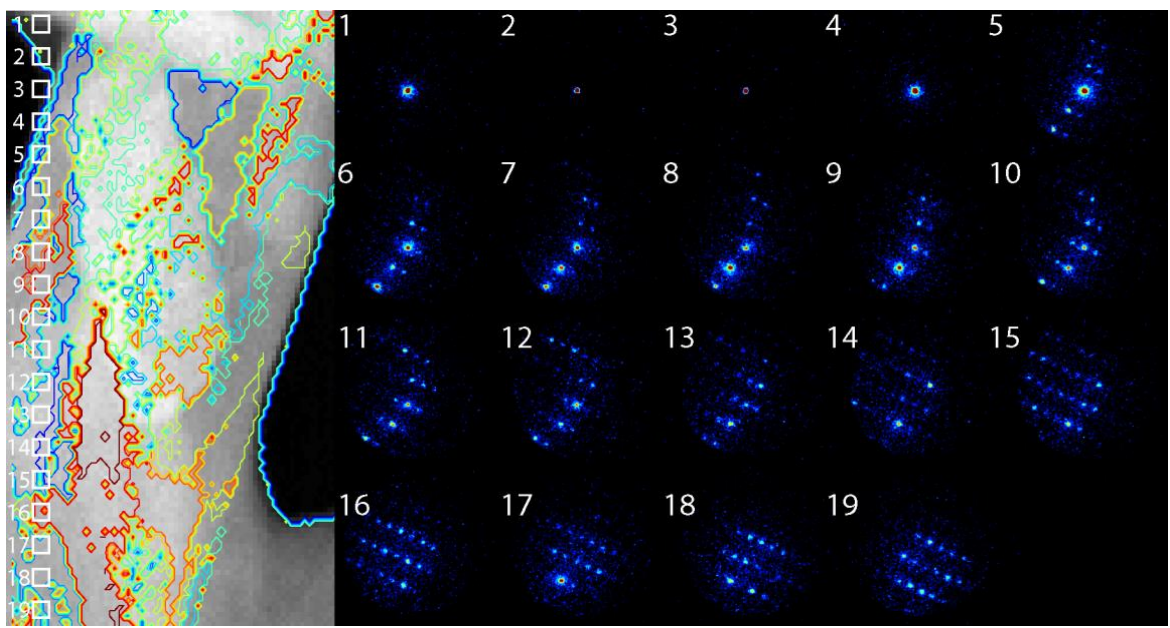

Supplementary Figure 10. **Smooth variation of diffraction as a function of spatial location within a nanocrystal.** (**Left**) HAADF image of a QYNNQNFV nanocrystal with contour lines outlining the clusters presented in Fig. 2(**b**). The numbered boxes indicate the region of the crystal that diffraction patterns (**right**) were acquired from. These patterns represent a 3x3 spatial average of diffraction data within the boxed region.

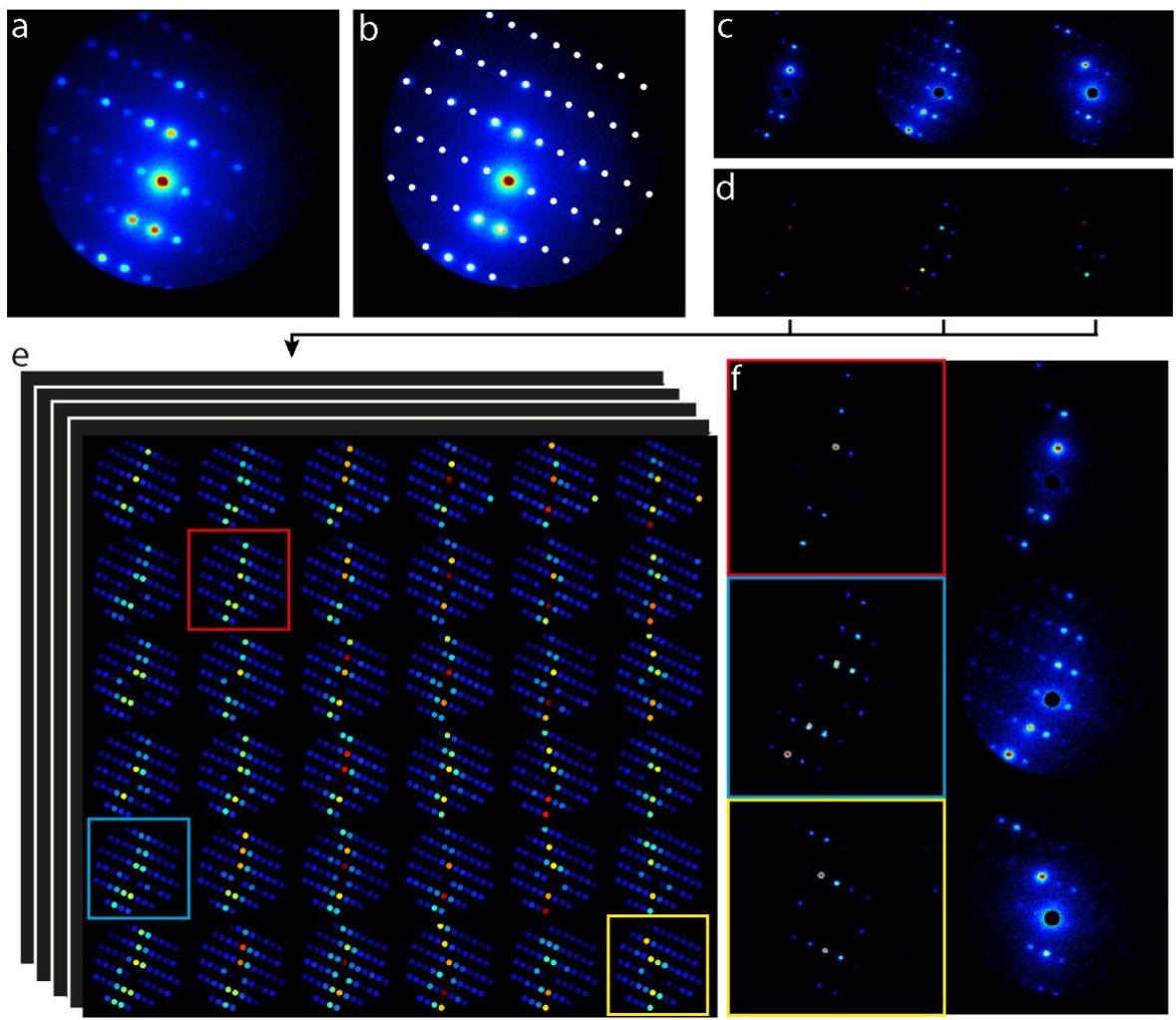

Supplementary Figure 11. **Orientation assignment of nanodiffraction cluster averages by library matching.** (**a**) Summed intensity of all patterns within a 4DSTEM scan. (**b**) Identification of all possible Bragg peaks that could be excited within the bounds of the HAADF detector. (**c**) Average patterns from individual crystalline regions. (**d**) Reduction of patterns to a list of intensities (represented by solid circles) and their respective *kxky* locations. (**e**) Library of simulated NBED patterns, best matches are highlighted by coloured boxes. (**f**) Results of library matching with best matched simulated patterns (left) and the corresponding cluster averages (right).

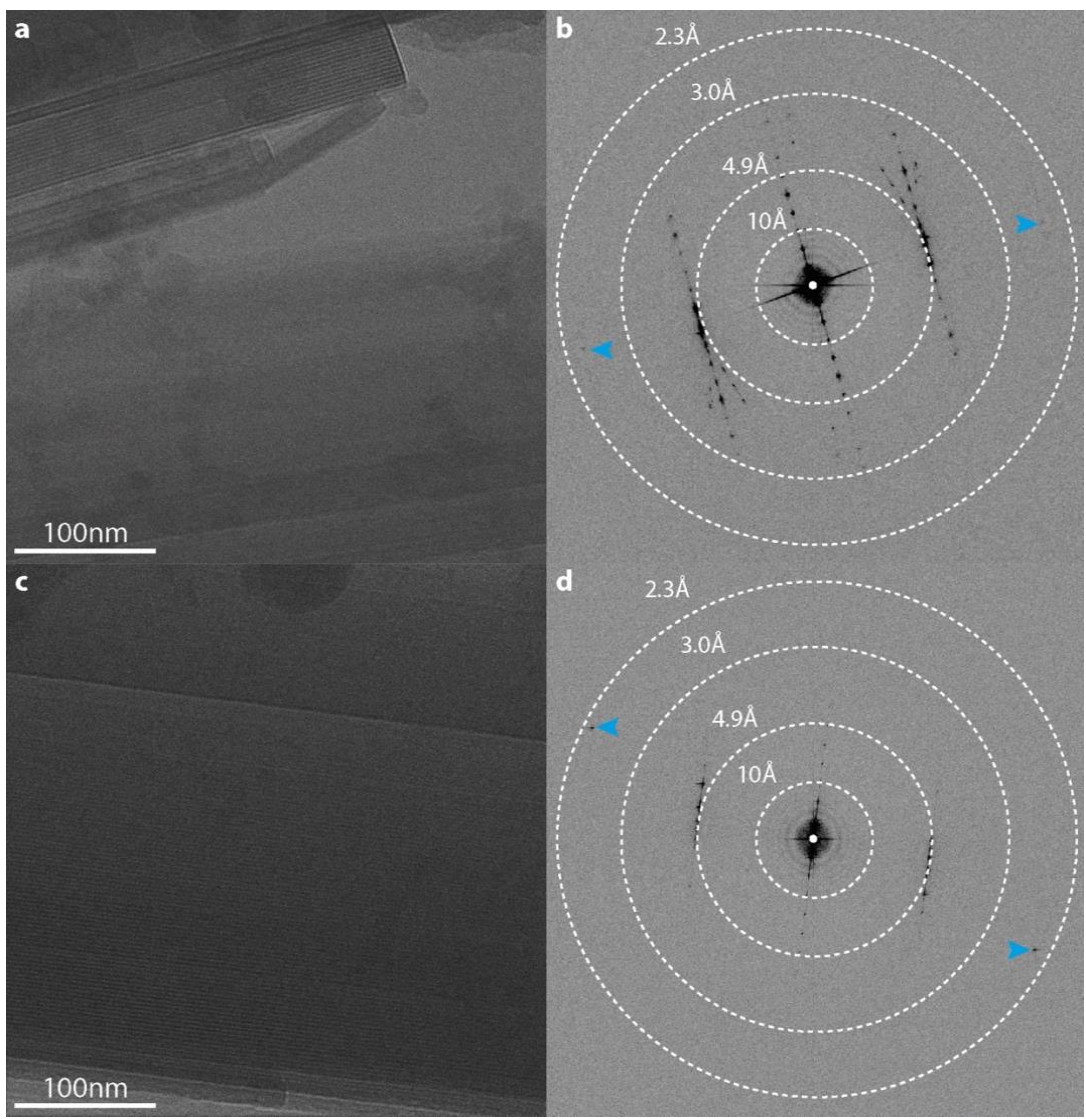

Supplementary Figure 12. **Crystal lattices observed in HRTEM images of crystal clusters and single nanocrystals.** (**a**) High resolution TEM image as shown in Fig. 3a. (**b**) Fourier transform of (**a**) with resolution rings. Blue arrows indicate highest resolution Bragg peaks at just less than 2.3Å resolution (**c**) High resolution TEM image as shown in Fig. 3d. (**d**) Fourier transform of (**c**) with resolution rings. Blue arrows indicate highest resolution Bragg peaks at just less than 2.3Å resolution.

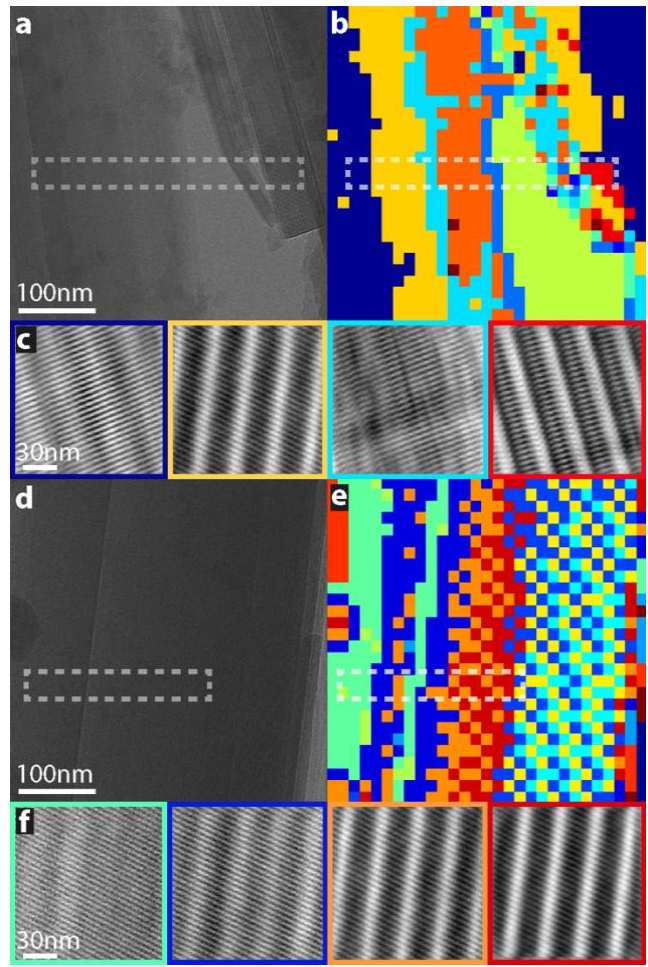

Supplementary Figure 13. **Mapping of crystalline regions within peptide nanocrystals by HRTEM.** (**a**) HRTEM image of a cluster of QYNNQNNFV nanocrystals. (**b**) Map of lattice changes present in the region imaged in (**a**). (**c**) Fourier filtered averages of regions within the dashed white box shown in (**a**) and (**b**). (**d**) HRTEM image of a single QYNNQNNFV nanocrystal. (**e**) Map of lattice changes present in the single crystal visible in (**a**). (**f**) Fourier filtered averages of regions in the dashed white box in (**d**) and (**e**).

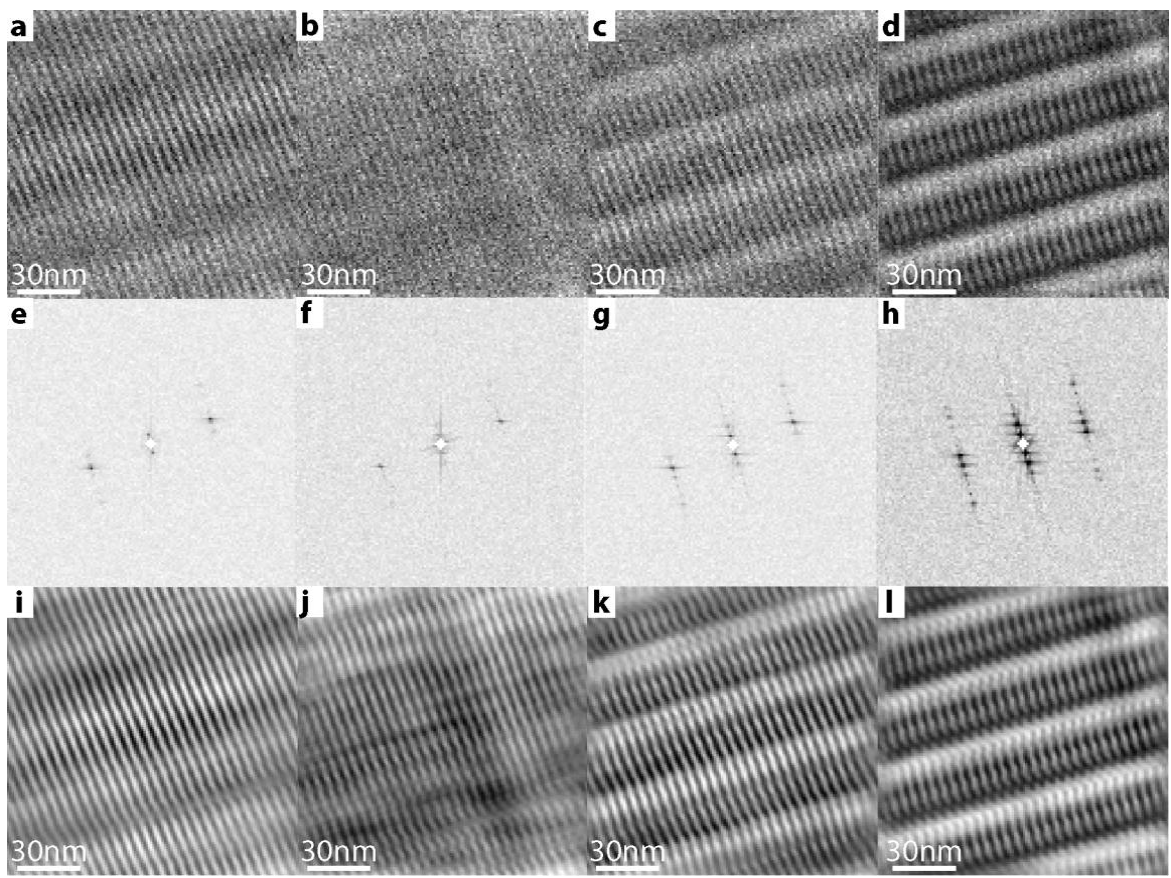

Supplementary Figure 14. **Fourier filtering of HRTEM cluster averages**. (**a - d**) Raw cluster averages before filtering from the image in Fig. 3a. (**e - h**) Fourier transforms of (**a - d**) showing the presence of Bragg peaks within the image transforms. (**i - l**) Cluster averages after Fourier filtering (as Fig. 3c).